\documentclass[cameraready]{Interspeech}
\usepackage{xurl}

\title{Is Natural Always Appropriate? Investigating Naturalness and Appropriateness Across Different Domains for TTS Evaluation}

\author[affiliation={1}, correspondingauthor]{Dominika}{Woszczyk}
\author[affiliation={2}]{Andreas}{Triantafyllopoulos}
\author[affiliation={3}]{Jura}{Miniota}
\author[affiliation={3}]{Éva}{Székely}
\author[affiliation={2,4}]{Bjoern}{Schuller}

\address{
    $^1$ Iconic, United Kingdom \\
    $^2$ Technische Universität München, Germany \\
    $^3$ KTH Royal Institute of Technology, Sweden \\
    $^4$ Imperial College London, United Kingdom
}

\email{dominika@iconicgames.ai}


\usepackage{comment}
\usepackage{tabularx}

\usepackage[activate]{microtype}
\begin{document}
\maketitle

\begin{abstract}
Text-to-speech (TTS) evaluation is an open challenge. While the primary target was ``naturalness," recent fidelity gains shifted focus toward ``appropriateness" and whether speech is correct for its context. In this work, we examine how perception changes when the expected downstream use varies. We measure the appropriateness and human-likeness of five SOTA TTS systems across five domains: AI assistant, reader, actor, animated character, and spontaneous speaker. Results show appropriateness varies across domains independently of naturalness. While systems shine at reading, expressive domains remain challenging, and optimizing for one can degrade others. Furthermore, naturalness scores tend to penalize stylized speech while rewarding spontaneity. Finally, our study also highlights blind spots in one-size-fits-all evaluation metrics across more expressive domains. We demonstrate that TTS performance is not ``solved" but depends on the target domain, requiring context-aware evaluation.
\end{abstract}

\section{Introduction}
Text-to-speech is fundamentally a one-to-many problem: the same sentence can be spoken in countless ways while remaining intelligible. Prosody, pacing, and delivery vary naturally depending on the situation, the speaker’s intent, and the audience. Consequently, good synthesis depends entirely on delivery and performance style. A voice perfectly suited for an audiobook may be inappropriate for an emergency hotline, a conversational agent, or an animated character~\cite{triantafyllopoulos2023overview, wagner2019speech}.

For a long time, naturalness has been the primary evaluation target in speech synthesis. However, naturalness is now widely recognized as an ill-defined and multi-dimensional concept~\cite{shirali2025multi, le2024limits}. Importantly, Mean Opinion Scores (MOS), the most common measure of naturalness, are neither stable nor absolute. Ratings depend on evaluation setup and listener expectations, and framing the intended use can change system rankings~\cite{le2024limits, perrotin2025refining, edlund2024assessing, shirali2023better}. As systems improve and reach high MOS values, differences that matter in real applications are not always reflected in a single naturalness score.

For this reason, several automatic metrics attempt to provide ``objective'' alternatives~\cite{cooper2024}, ranging from ASR-based intelligibility to spectral fidelity and distributional measures of prosodic similarity~\cite{minixhofer2024ttsds, minixhofer2025ttsds2}. In expressive TTS, evaluation often relies on emotional embedding distances or pitch correlation against a target sample or style~\cite{triantafyllopoulos2023overview}, while recent persona-based benchmarks assess instruction following and character consistency~\cite{jiang2025speechrole, shi2025speech, huang2025instructttseval}.
More recently, persona-based benchmarks assess instruction following, role-play ability, or character consistency~\cite{jiang2025speechrole, shi2025speech, huang2025instructttseval}. 

While these methods extend evaluation beyond simple MOS, they still provide only partial views. Many focus on global similarity or reconstruction accuracy rather than contextual suitability, and some approaches, such as LLM-as-judge~\cite{manku2025emergentttseval, shi2025speech}, raise important questions regarding their reliability. Furthermore, current benchmarks prioritize linguistic ``stress tests'' (e.g., difficult pronunciations) over functional performance across spontaneous or expressive domains~\cite{lajszczak2024base, manku2025emergentttseval}. Crucially, given that these ``task-agnostic'' scores ignore situational needs, a model with high accuracy may still be perceived as inappropriate for its intended real-world use.

In parallel, as ``naturalness'' is being increasingly questioned by the research community, several works argue that evaluation should consider ``appropriateness'' -- that is, whether speech fits its intended communicative purpose~\cite{pandey25_ssw,triantafyllopoulos2023overview}. This shift moves the focus from ``human-likeness'' to evaluating how suited the generated response is given the target task. 
However, the impact of the target domain on appropriateness judgments has not been systematically studied. Existing work often compares read versus spontaneous speech~\cite{dall2014rating}, or examines how accented or disordered speech affects perceived naturalness~\cite{mackey1997effect}. Appropriateness is also partially explored in terms of style transfer or role-play settings, where the goal is to reproduce a specific persona~\cite{shi2025speech, lee2025p2va}. 
Previous work thus leaves an important gap that needs to be investigated -- how the same utterance is judged when framed for different target tasks that require different operationalizations of expressivity.


\textbf{Our Study} We contribute to this ongoing discussion by investigating how perceived appropriateness for both synthetic and human speech changes across five domains: read speech, acted speech, spontaneous interaction, conversational assistant, and animated character. This study examines how framing the intended use affects listener expectations, tolerance, and whether naturalness, which we frame as human-likeness, reliably predicts suitability. Our contributions are as follows:
\begin{itemize}

\item \textbf{Cross-Domain Perceptual Analysis}
We systematically measure how perceived naturalness and appropriateness change across various settings for different TTS.

\item \textbf{Human-Likeness Paradox Exploration}
We analyze the relationship between human-likeness and domain suitability, and show that while they may align in certain contexts, in others human-likeness may not predict appropriateness and vice-versa.

\item \textbf{Domain-Specific Evaluation \& Metrics Profiling}
We show that TTS systems perform differently across different domains, and that common metrics are not universally indicative of appropriateness.

\end{itemize}

\section{Methodology}

\subsection{Perception Study Design}

To evaluate the perception of appropriatness of synthetic speech across domains, we conduct a perceptual study. We design the listening test using a \texttt{Gradio} interface hosted on Prolific 
\cite{HomeProl50:online}\footnote{https://github.com/domiwk/domain-aware-tts-eval}\footnote{Study demo page available at https://researcht81.github.io/unconvincing-human}. We recruited 150 native English speakers (95\%
approval rating) via Prolific, split into 6 sessions of 25 participants.
We curated 30 sentences with ground truth (GT) samples and synthesized them across 5 TTS systems. Using a Latin Square design, 180 samples ($30 \times (5+1)$) were distributed across six sessions. Each participant evaluated all 30 sentences and all systems without repetition, ensuring independent judgment while anchoring across TTS profiles. Two attention checks were included to filter bots and inattentive participants. 



 
\vspace{3pt} \noindent \textbf{Rating Task} Participants rated both human-likeness and appropriateness (for each persona) on a 5-point Likert scale. After initial pilot studies observations, we use the term ``convincingness'' instead of appropriateness to improve conceptual clarity and reduce bias, and naturalness as ``human-likeness'' to make it a more defined concept.

\vspace{3pt} \noindent \textbf{Stimuli Curation} We manually curated speech and text pairs spanning four speech task families (examples shown in Table~\ref{tab:example-sentences-by-dataset}). We selected samples from task-related datasets and extracted corresponding samples as ground truth:
\begin{itemize}
  \item \textbf{Narration}: We extracted quotation-style narration excerpts curated from LibriQuote~\cite{michel2025libriquote} read speech (narration-only subset), designed to probe descriptive statements.
  \item \textbf{Spontaneous conversational}: We selected conversational sentences from MSP-Podcast~\cite{lotfian2017building} to stress-test informal phrasing, disfluencies, and natural conversational prosody in realistic audio conditions.
  \item \textbf{Affect conversational}: We curated acted dialogue using MELD \cite{poria-etal-2019-meld}
   and AnimeVox~\cite{taresh1826:online}, to probe emotional expressivity and as GT for acted and animated character.
  \item \textbf{Inform}: We generated 6 sentences using \texttt{gemini-3-pro} that are informative and conversational across different statement types (statement, interrogation) and length, and manually checked them. We generated a proxy for the AI assistant GT using \texttt{elevenlab-v3} and the voice \textit{Katie X}. 
\end{itemize}

In order to mitigate participants confusing
the content appropriateness with the delivery and get a diverse coverage of emotions for the conversational sets, we first ran an LLM pass to label the sentences for plausible emotions from the text and for appropriateness for each given persona. We then manually selected sentences that are less idiomatic for the source persona and for the conversational tasks to span a total of 6 emotions per stimuli set (anger, fear, sadness, disgust, neutral, joy).


\begin{table}[!h]
\centering
\small
\resizebox{\columnwidth}{!}{%
\begin{tabular}{@{} l p{0.78\columnwidth} @{}}
\toprule
\textbf{Speech Task} & \textbf{Example Sentence} \\
\midrule
Inform & Your meeting starts in 30 minutes. \\
Affect Conversational & Ew! What is that? Something exploded!\\
Spontaneous & Oh, um, I can't. I mean, I don't know the festival circuit and all that. \\
Narration & But that night Dorothy could not sleep. The excitement perhaps, or was it fear? \\
\bottomrule
\end{tabular}%
}
\caption{Example sentences per speech task.}
\label{tab:example-sentences-by-dataset}
\end{table}

\vspace{3pt} \noindent \textbf{TTS Systems}
We identified high-quality TTS systems from the Emergent TTS benchmark~\cite{manku2025emergentttseval} with low WER ($\leq$0.13), and $\geq$ 20\% win rate overall, and that cover different stylistic archetypes across the expressivity and spontaneity axis. To reduce voice-preference bias, we selected female voices with similar timbre.

\begin{itemize}
\item \textbf{Kokoro} (\textit{af\_heart}): A lightweight 82M parameter StyleTTS 2 model. Provides high-quality speech but is prosodically consistent, with low spontaneity.

\item \textbf{Gemini TTS} (\textit{Flash 2.5, Despina}): Google's flagship multimodal model and first on the Emergent TTS leaderboard. It delivers highly expressive and stylized speech.

\item \textbf{Kyutai-TTS} (\textit{1.6B, p037}): Built on the Moshi audio-to-audio framework. It was trained on raw conversational data and captures high spontaneity and natural disfluencies.

\item \textbf{GPT-4o-mini-tts} (\textit{Coral}):  A high-quality commercial TTS with a balanced profile with medium--high spontaneity and moderate expressivity.

\item \textbf{ElevenLabs} (\textit{multilingual\_v2, Bella}): A state-of-the-art commercial TTS model that is designed for professional delivery. We picked the default voice (``Bella'') for conversational speech with medium spontaneity and moderate expressivity.  

\end{itemize}

\subsection{Acoustic Features and Automatic Metrics}

\vspace{3pt}\noindent\textbf{Acoustic Features} To investigate the characteristics linked with appropriateness within each domain and identify the preferred acoustic profiles, we analyzed features spanning \textit{rhythm} (articulation rate sd, speech rate, nPVI), \textit{expressivity} (f0 range semitones, percentiles, RMSE sd, arousal, valence), and \textit{voice quality} (jitter, shimmer, H1-H2, alpha ratio, CPPS).\footnote{Computed with  \texttt{praat-parlsemouth} and eGeMAPSv02~\cite{eyben2015geneva} through 
openSMILE~\cite{eyben2010opensmile} and  \texttt{wavlm-large-msp-podcast-emotion-dim}~\cite{feng2025vox} for valence and arousal. }

\vspace{3pt}\noindent\textbf{Automatic Metrics} To investigate the effectiveness of automatic evaluation, we analyzed metrics spanning \textit{quality estimation} (UTMOSv2~\cite{baba2024utmosv2}, DNSMOS~\cite{reddy2021dnsmos}, Squim~\cite{kumar2023torchaudio}, PESQ~\cite{rix2001perceptual}, MCD~\cite{mcd}, STOI~\cite{stoi}), \textit{prosodic distance} (f0 correlation computed with SwiftF0~\cite{nieradzik2025swiftf0}, AutoPCP~\cite{barrault2023seamless}, WavLM~\cite{chen2022wavlm}), \textit{style} (AudioBox CE/CU/PQ~\cite{vyas2023audiobox}), \textit{intelligibility} (WER, measured with nvidia/parakeet-tdt-0.6b-v2~\cite{nvidiapa29:online}), and \textit{diversity} (DS-WED~\cite{yang2025measuring}).

\section{Results}
\begin{figure*}[h]
    \centering
\includegraphics[width=0.92\textwidth]{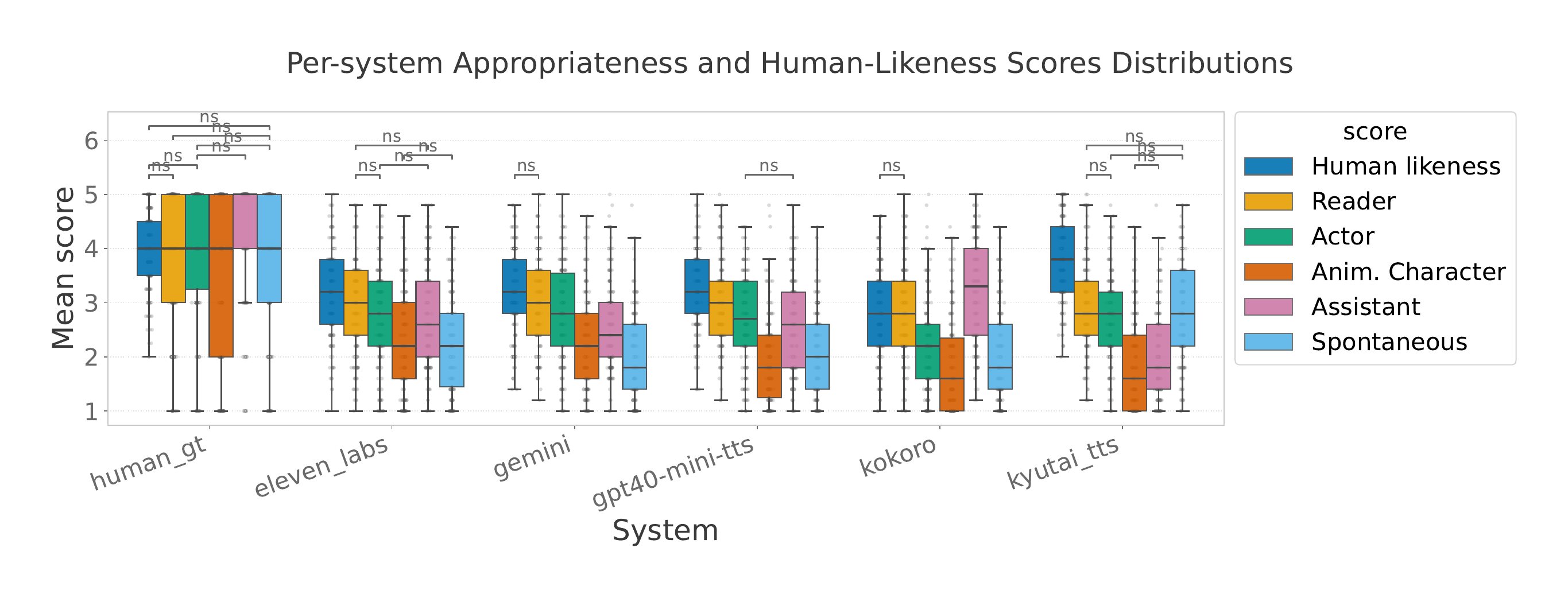}
    \caption{Appropriatness and human-likeness score for each TTS across the 5 personas across all speech tasks, averaged across sentences per participant per session. For ground truth, we report the scores only for dataset-matched sentences as the upper anchor. We mark non-significant pairs with `ns` with the Wilcoxon paired test (with Holm–Bonferroni corrections) and p-value$\leq$0.05.}
    \label{fig:app_per_tts_overall}
\end{figure*}

\subsection{Appropriateness Across Domains}

Figure~\ref{fig:app_per_tts_overall} shows the distributed score for each persona across the TTS and all tasks. The results show that while most systems achieve high appropriateness scores for reading or AI assistant roles, for some contexts like spontaneous conversation, actor, and animated character personas remain more challenging. 

Specifically, Kokoro performs well for reading and assistant tasks but scores poorly on conversational tasks. Conversely, Kyutai-TTS is perceived as highly appropriate for spontaneous conversation but has low scores for the AI assistant and animated character personas. Systems like Eleven Labs, Gemini, and GPT-4o-mini-TTS score highly on the acting domain but fail to achieve high scores in the spontaneous conversation.  We report an overall Krippendorff's $\alpha$ of 0.2 for TTS and 0.44 for GT samples, indicating a relatively low level of agreement.

These results show that a TTS system that is deemed appropriate for one domain is not necessarily suitable for another. Kokoro's narrow profile and peaks in assistant and reading roles, despite lower naturalness, suggest listeners may actually expect AI assistants to sound somewhat robotic.
Meanwhile, Kyutai-TTS, while offering a broader profile, excels in spontaneous speech but sounds too raw for an assistant. Finally, the low inter-rater agreement suggests appropriateness is highly subjective, heavily influenced by individual listener expectations.

\subsection{Human-likeness and Appropriateness across Domains}

\begin{table}[!h]
    \centering
    \caption{Correlation between Human-likeness and Appropriateness across Domains.}
    \label{tab:correlation}
    \small
    \resizebox{\columnwidth}{!}{ 
    \begin{tabular}{lccccc}
        \toprule
        \textbf{Domain} & Spontaneous & Actor & Reader & Anim. Character & Assistant \\
        \midrule
        \textbf{Spearman $\rho$} & 0.4021 & 0.4705 & 0.3757 & 0.0821 & -0.4438 \\
        \bottomrule
    \end{tabular}
    }
\end{table}

Table \ref{tab:correlation} shows the correlation between human-likeness and style scores at the sentence level. This indicates how much the perceived naturalness aligns with the specific style requirements. We observe positive correlations for Actor, Spontaneous, and Reader domains, whereas correlations are near-zero for Animated Character and negative for Assistant. We hypothesize that while systems balancing prosodic naturalness with expressivity achieve higher appropriateness, those like Kokoro, though deemed highly appropriate for the Assistant role, yield lower naturalness scores. Conversely, Kyutai-TTS demonstrates that high naturalness does not guarantee appropriateness for Animated or Assistant voices, leading to the observed decoupling of these metrics in those domains.

\subsection{Impact of Speech Tasks on Naturalness}

 \begin{figure}[!h]
    \centering
    \includegraphics[width=0.9\columnwidth]{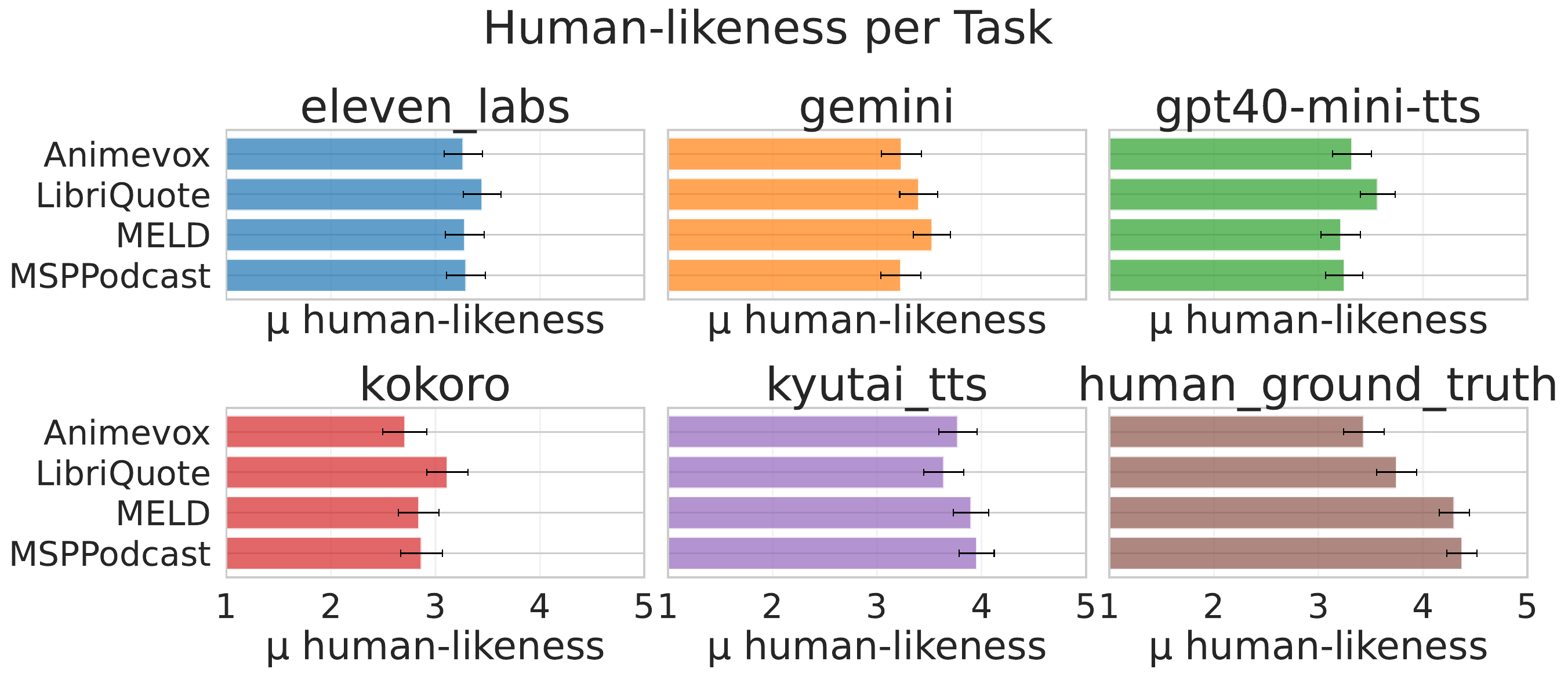}
    \caption{Human-likeness mean scores for systems across different speech tasks.}
    \label{fig:hl_per_task}
\end{figure}


Figure \ref{fig:hl_per_task} illustrates the task's impact on human-likeness scores. Interestingly, even GT ratings varied across domains, with conversational samples from MELD and MSP Podcast preferred over LibriQuote and Animevox. Manual inspection suggests these lower scores likely come from the Irish accent or the more mature voice of the reader, which participants may have perceived as less natural. This finding aligns with previous studies on non-standard speech. Similarly, the lower scores for Animevox set, but not for the MELD set, suggest that highly stylized delivery is penalized regardless of the task.

On the other hand, TTS naturalness scores may reflect a model's ability to handle specific styles rather than the task's inherent difficulty. For instance, while narration is often considered easier to synthesize, Kyutai-TTS performed more naturally on conversational tasks than on reading. Interestingly, a similar analysis of mean appropriateness scores across speech tasks revealed no considerable
differences.

\subsection{Appropriateness and Acoustic Features}

 \begin{figure}[!h]
    \centering
    \includegraphics[width=0.84\columnwidth]{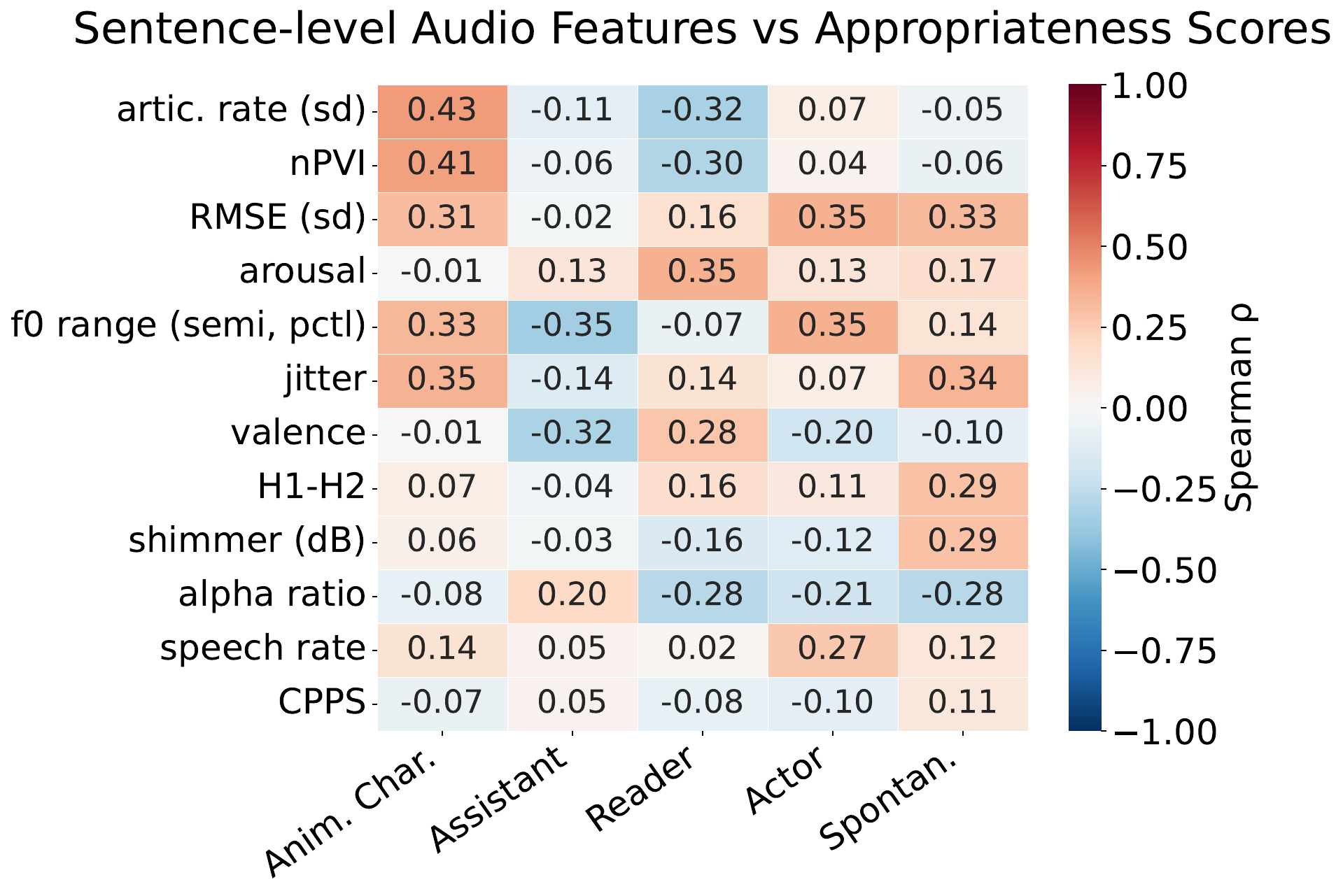}
    \caption{Appropriateness-acoustic features correlation on sentence level across all speech tasks.}
    \label{fig:heatmap_acc_corr}
\end{figure}

To understand what features are more suitable for each domain, we analyze the Spearman correlations between sentence-level appropriateness scores and acoustic features.

In Figure~\ref{fig:heatmap_acc_corr}, we observe that Animated Character correlates most strongly with articulation rate variability ($\rho=0.43$) and nPVI ($\rho=0.41$). For this domain, ``good" synthesis seems to require significant fluctuations in pacing and rhythm. In contrast, the Reader domain shows negative correlations with these same features ($\rho \approx -0.30$), suggesting that listeners prefer a more steady, controlled rhythmic delivery for long-form reading. For the Assistant, we notice a preference for stability and neutrality. This is reflected in the negative correlation with f0 range ($\rho=-0.35$) and valence ($\rho=-0.32$), indicating that overly expressive or emotionally charged speech is deemed less appropriate for an AI companion. Meanwhile, high-expressivity domains like Actor and Spontaneous speech show positive correlation to RMSE (sd $\approx 0.35$) and voice quality. Specifically, Spontaneous speech is the only domain to show notable positive correlations with features like jitter ($\rho=0.34$) and creakiness (H1-H2, $\rho=0.29$), suggesting that these ``human'' imperfections are actually expected in casual interaction. Finally, we see that spectral tilt (alpha ratio) has a slight positive correlation with Assistant appropriateness ($\rho=0.20$); it is penalized in Spontaneous ($\rho=-0.28$) and Actor ($\rho=-0.21$) contexts.

These results show that appropriateness is not only linked with rhythmic features (more or less expressive or energetic) but also with the quality of the voice itself. While a TTS model can be trained to adapt its rhythmic properties given the linguistic context, voice qualities are harder to adapt at test time. This suggests that both the design of the voice and the TTS system need to take into account the target delivery style.

\subsection{Appropriateness and Automatic Metrics}

In Figure 4, we examine the correlation between sentence-level appropriateness and various automatic metrics. We observe that common quality estimators, UTMOS and DNSMOS, show an important 
negative correlation with appropriateness for Actor ($\rho \leq -0.41$) and Spontaneous ($\rho \leq -0.47$) styles. This suggests that these metrics penalize the naturalistic disfluencies and high-dynamic range essential for expressive or conversational speech. Conversely, Assistant correlates positively with these MOS predictors ($\rho \approx 0.35$), indicating that for traditional high-quality audio, these metrics remain a valid proxy for appropriateness.

Furthermore, we notice that embedding-based metrics like AudioBox and AutoPCP are effective for styles like Assistant and Reader ($\rho \geq 0.36$), yet they fail to capture the nuances of human-likeness, where they show stronger inverse correlations ($\rho \approx -0.50$). It is also noteworthy that WavLM distance is negatively correlated with Assistant($\rho = -0.72$), but positively correlated with human-likeness, indicating that WavLM embeddings are more representative of neutral speech. Finally, DS-WED is a more robust metric for diversity, whereas PESQ and f0 correlation remain largely uninformative for predicting perceived appropriateness.

 \begin{figure}[!h]
    \centering
    \includegraphics[width=0.85\columnwidth]{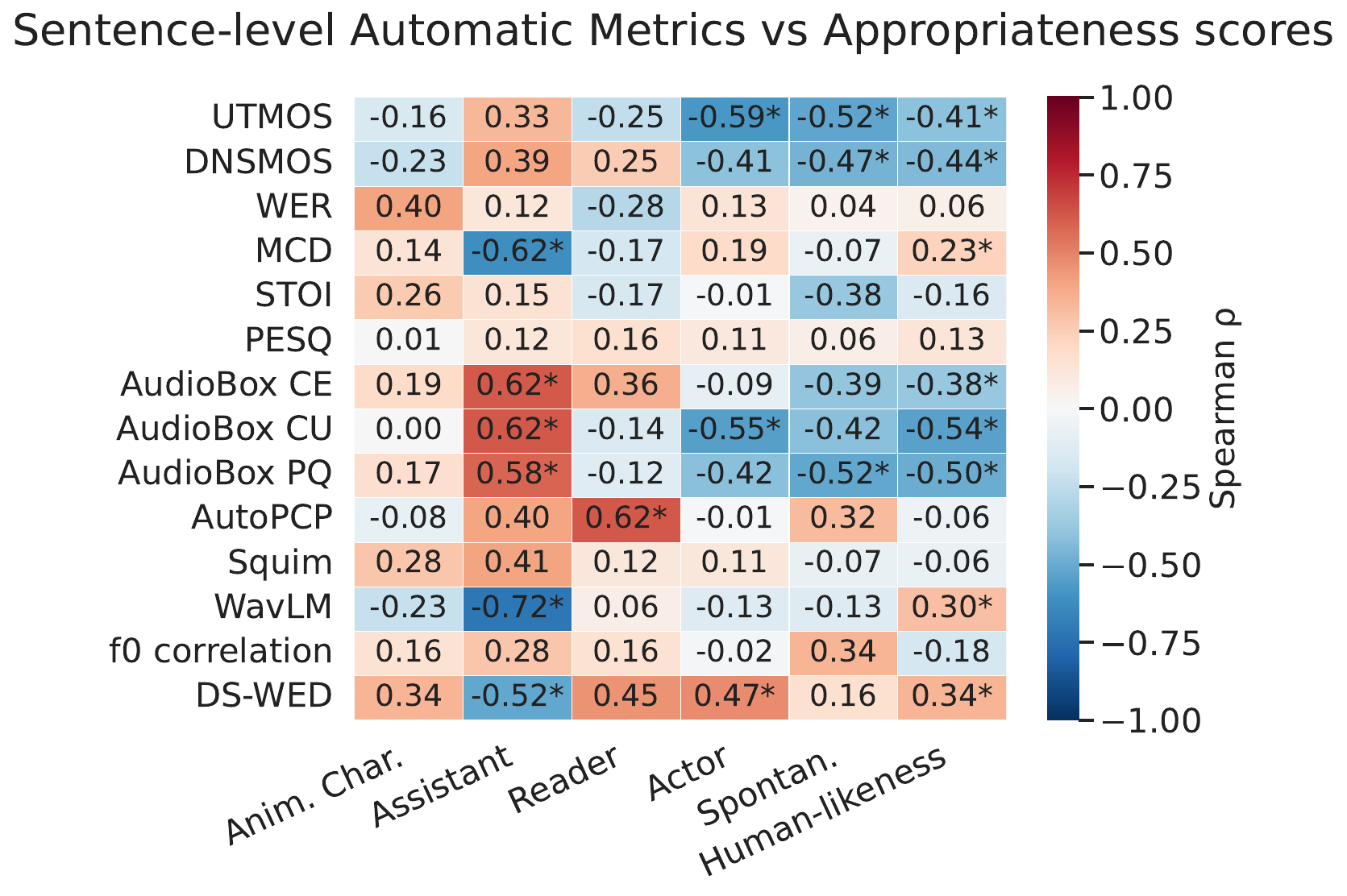}
    \caption{Appropriateness-automatic metrics correlations on sentence level across all speech tasks. * indicates significance.}
    \label{fig:heatmap_spearman}
\end{figure}

\section{Discussion \& Limitations}

This study demonstrates that speech evaluation is inherently domain-dependent. Listener judgments for the same utterance shift based on the framed scenario, showing that perceived quality depends as much on situational expectations as the acoustic signal itself. Our findings confirm that human-likeness and appropriateness are distinct dimensions. Human-likeness fails to reliably predict cross-domain suitability and tends to favor spontaneous over stylized speech. Because even highly human-like voices can feel inappropriate for specific tasks, evaluation must prioritize metrics aligned with the intended use. Furthermore, even flexible systems struggle to capture the domain-specific voice qualities required across diverse applications. We also observe that different TTS systems favor specific domains rather than uniform performance. Systems likely reflect their training data and optimization targets, supporting the idea that TTS systems and chosen voices are not a one-size-fits-all solution.

A limitation of this study is the use of isolated sentences. Real-world speech involves dialogue and emotional progression and extending this evaluation to multi-turn contexts is a natural next step. Additionally, we did not explore how a speaker’s perceived gender, age, or socioeconomic background influences perception. Investigating how these social identities and simulated roles shape listener perception remains an important avenue for future research.

\section{Conclusion}

In light of the recent improvement in TTS performance, the question of quantifying progress has become more pertinent than ever.
Most works focus on \emph{naturalness}, a term which is usually equated with ``appearing human-like'', in colloquial terms.
Yet, our listening experiments show that the \emph{appropriateness} of a response is context and application-dependent; there is
no one-size-fits-all approach that fits all intended uses of a TTS system.
Rather, the correlation between human-likeness differs across different applications.
Moreover, state-of-the-art TTS systems do not score equally well across all evaluated dimensions, but rather show a differentiated, domain-specific profile.
Our work illustrates how evaluations of TTS systems should be multi-dimensional and contextualized. Rather than asking if an utterance ``sounds like a human'', we should be asking whether an utterance ``sounds right''.
Future work could further investigate how appropriateness is manifested across different application areas and, crucially, how it can allow us to better gauge performance and iteratively define future TTS systems.

\section{Generative AI Use Disclosure}
Generative AI was used to edit and polish drafts made by the authors. Any generated content was reviewed and edited by authors who maintain full responsibility for the final content.

\bibliographystyle{IEEEtran}
\bibliography{lib.bib}

\end{document}